\documentclass[conference]{IEEEtran}

\usepackage{enumitem}
\usepackage{balance} 
\usepackage{amsmath,amssymb} 
\usepackage{graphicx}
\usepackage{subcaption} 
\usepackage{booktabs}
\usepackage{multirow,xcolor} 
\usepackage{algorithm}
\usepackage{algpseudocode}
\usepackage[hyphens]{url}
\usepackage[hyphenbreaks]{breakurl}

\begin{document}

\title{A Survey on Amazon Alexa Attack Surfaces}

\author{
  \IEEEauthorblockN{Yanyan Li\IEEEauthorrefmark{2},
    Sara Kim\IEEEauthorrefmark{3},
    Eric Sy\IEEEauthorrefmark{3}}
  \IEEEauthorblockA{
    Department of Computer Science and Information Systems\\
    California State University San Marcos,
    San Marcos, California 92078\\
    Email: \IEEEauthorrefmark{2}yali@csusm.edu,
    \IEEEauthorrefmark{3}\{kim106, sy004\}@cougars.csusm.edu}
}

\maketitle

\begin{abstract}
  Since being launched in 2014, Alexa, Amazon’s versatile
  cloud-based voice service, is now active in over 100 million
  households worldwide \cite{bohn2019verge}.
  Alexa’s user-friendly, personalized vocal experience offers
  customers a more natural way of interacting with cutting-edge
  technology by allowing the ability to directly dictate commands
  to the assistant. Now in the present year, the Alexa service is
  more accessible than ever, available on hundreds of millions of
  devices from not only Amazon but third-party device manufacturers.
  Unfortunately, that success has also been the source of concern
  and controversy. The success of Alexa is based on its effortless
  usability, but in turn, that has led to a lack of sufficient
  security. This paper surveys various attacks against Amazon
  Alexa ecosystem including attacks against the frontend voice
  capturing and the cloud backend voice command recognition and
  processing. Overall, we have identified six attack surfaces
  covering the lifecycle of Alexa voice
  interaction that spans several stages including voice data
  collection, transmission, processing and storage.
  We also discuss
  the potential mitigation solutions for each attack surface to
  better improve Alexa or other voice assistants in terms of
  security and privacy.
\end{abstract}

\begin{IEEEkeywords}
  Alexa skills, Amazon Alexa, attack surfaces, Echo, Internet of
  Things, privacy, security, voice hacking
\end{IEEEkeywords}

\section{Introduction}
\label{sec:introduction}

Imagine we are locked inside a room that contains a door, a window,
and a vent. Outside the door are one thousand ninjas all trying to
attack us. What are the different ways the ninjas can access our
room? The answer: the door, the window, and the vent. These access
points are the ``attack surface'' of the room. An attack surface,
for a software environment, can formally be defined as the sum of
different points (also known as ``attack vectors'') where an
unauthorized user (an ``attacker'') can try to enter or extract
data from an environment. Keeping an attack surface minimized is a
crucial security measure for any software \cite{manadhata2004measuring}.
The Amazon Alexa Voice assistant is a very new technology. This
innovative new software is set up to be not just a personal device
but a natural extension of the home environment. As a result, Alexa
not only has access to typical customer data used by smartphones
and laptops such as messages and schedules, but also has access and
control over house locks, personal shopping lists, voice recordings,
conversations, customer voice profiles and etc.

In April of 2019, Amazon disclosed the shocking extent of private
user data that Alexa carries, admitting that Alexa not only
continuously listens in on customer conversations but records and
saves that data \cite{day2019time}. Shockingly, Amazon has admitted
to employing full teams of people to listen in on users’
conversations with Alexa in order to improve Alexa’s perceived
weakness with foreign languages, regional expressions, and slang,
storing potentially very sensitive and private information that
could be at risk for exposure \cite{day2019time}. Furthermore, with
the inclusion of more ``smart home'' devices such as the ``smart
locks'' (gate codes, key codes) and security cameras (live footage
of customer homes) becoming integrated into the home environment
and by extension, Alexa, Alexa has gained more access to customer
information than any other contemporary software in the shortest
amount of time.

In addition, comparing to typing or touching based interactions,
voice based interaction eliminates the need of finger touch.
While this new avenue of interaction may be convenient for people
in certain scenarios, it also introduces new issues. Voices from
TV and radio signals, replayed voices that mimic a person's live
voice, and even some inaudible sounds could be picked up by voice
assistants \cite{yuan2018all,pradhan2019combating,zhang2017dolphinattack}.
Moreover, voice hacking or voice spoofing is becoming a new
phenomenon where attackers hijack an individual’s unique voiceprint
or voice profile in order to steal his or her identity — a problem
that has become particularly bad in an era where speech-controlled,
voice-activated products are common \cite{johnson2018entre}.

Voice based attacks can happen not only in the frontend during
voice transmission from users to voice assistants, but also in the
backend for queried websites such as Wikipedia.
An example of such an attack is the infamous ``Burger King Advert
Sabotage'' \cite{wakefield2017bbc}. In 2017, Burger King released
an advert that was designed to activate Google Home smart speakers
and Android phones to describe Whopper burgers. The Wikipedia page
for the Whopper had been maliciously edited as the ``worst hamburger
product" and added cyanide to the list of ingredients. This resulted
in false information being spoken from users' voice assistants.

\begin{figure*}[htbp]
  \centering
  \includegraphics[width=.8\textwidth]{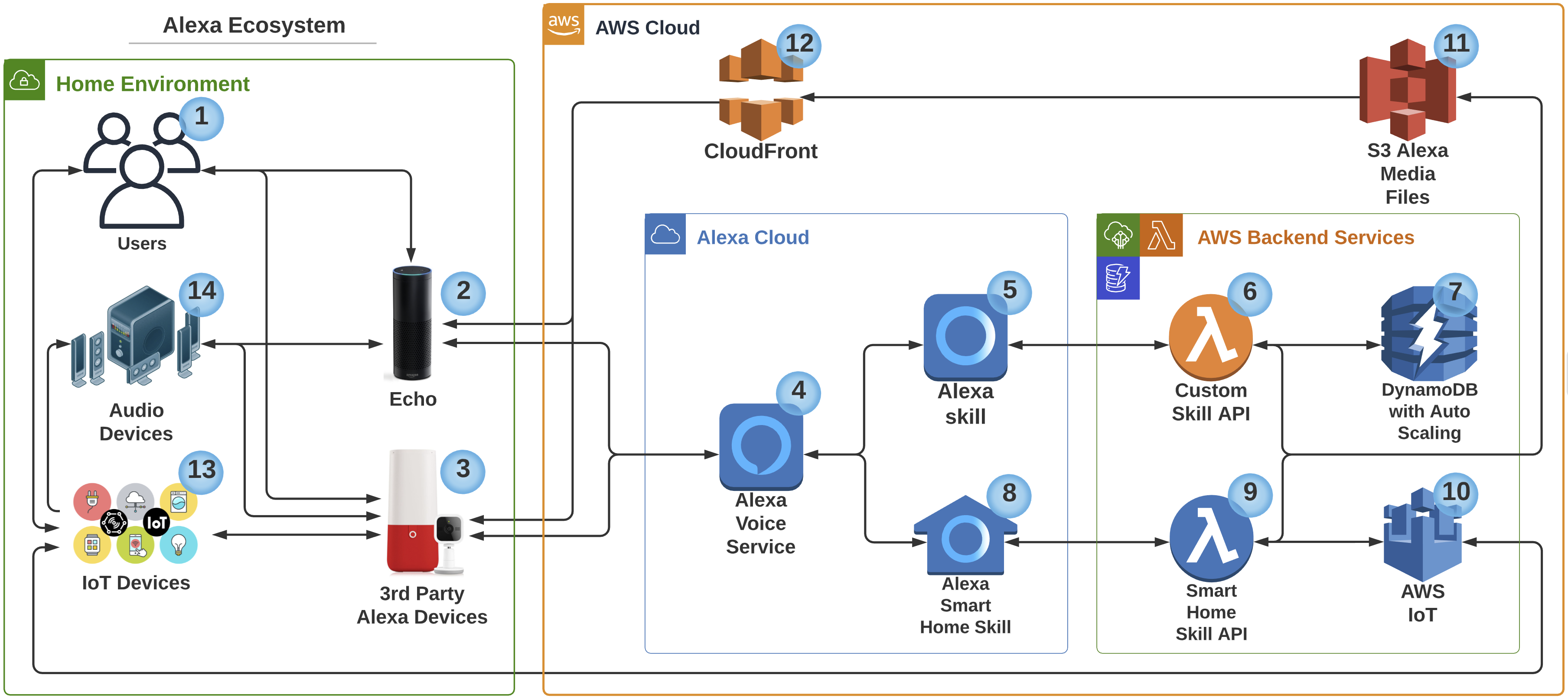}
  \caption{Overview of Alexa Ecosystem, consisting of the Home Environment, AWS cloud and its components}
  \label{fig:alexa-ecosystem}
\end{figure*}

Users interaction with voice assistants are not limited to asking
fact-based questions, checking the weather or playing music; but
could also be requesting a Uber ride, placing an Amazon order,
unlocking the front door and etc. In both situations, user voice
recognition and functions to process user request (e.g., pull data
from Wikipedia, send user requests to Uber or smart lock) are needed
in the cloud backend, but the latter ones are more critical as
the transcribed user voice command has total control over a user's
Uber/Amazon account and home security system.
If the voice recognition service is vulnerable (voices get
misinterpreted as a different command) or backend functions running
in the cloud get attacked, users' identities could be stolen,
accounts could be compromised, and home security could be at risk
\cite{barda2020checkpoint}.
The goal of this paper is to study Amazon Alexa attack surfaces to
shed light on building a more secure smart voice assistant system.

The main contributions of this work are as follow:
\begin{itemize}
  \item We presented an overview of Amazon Alexa ecosystem, its
        major components and their functionalities.
  \item We systematically studied the attacks against Amazon Alexa
        ecosystem and categorized those attacks into six attack surfaces,
        which are attacks in voice capturing, voice traffic transmission,
        Alexa voice recognition, Alexa skill invocation, Lambda functions
        and Amazon S3 bucket.
  \item We discussed the potential mitigation solutions to prevent
        against those identified attacks.
\end{itemize}

This paper is organized as follows. First, we present an overview
of Alexa system in Section~\ref{sec:alexa-system}. Then, we detail
its attack surfaces in Section~\ref{sec:attack-surfaces}. Finally,
we discuss the mitigation in Section~\ref{sec:mitigations}
and conclude the paper in Section~\ref{sec:conclusion}.

\section{Background/Alexa Ecosystem}
\label{sec:alexa-system}

The Alexa system consists of various physical input components that allow customers to interact with the system, referred to as “Echo” devices, and cloud components which include “smart” Speech Language Understanding (SLU) backend functionalities such as Automatic Speech Recognition (ASR), Natural Language Understanding (NLU), Text-to-Speech (TTS) conversation and Response \cite{na2019alexa}. ASR is a technology that allows computer interfaces to be communicated with in a way that resembles normal human conversation \cite{zajechowski2019usability}. Along the same vein, NLU is the comprehension by computers of the structure and meaning of human language and TTS is the ability to accurately transcribe human speech \cite{na2019alexa}. For the Alexa system, all these functionalities are used in order to process the best “response” to the customer’s interaction.

Many responses to customer interactions are provided directly from Alexa. However, responses can also be provided by third party services through “skills”. Skills are voice-driven applications that can be added to Alexa to extend functionality and personalize the user experience. Skills are launched and executed through the following interaction model \cite{kumar2018skill}:

\begin{enumerate}
  \item If a user says the wake word, “Alexa”, the Echo device will
        be activated and listen for user voice command.
  \item That command is sent to the cloud where it is processed via
        ASR and NLU and transcribed using TTS.
  \item A JavaScript Object Notation (JSON) request is sent to the
        skill's Lambda function, processing user intent.
  \item Once the request is processed, the Lambda function sends a
        JSON response to the Alexa voice service.
  \item The Alexa voice service receives the JSON response and
        converts the output text to an audio file.
  \item Echo device receives the audio response from the cloud
        and plays the audio via the built-in speaker.
\end{enumerate}

Skills, voice applications that process users requests, have become the new target for attackers due to their capabilities of accessing user voice commands. Malicious skills have been seen to steal user data or manipulate user voice command to execute tasks deviating from users' intentions.

The Amazon Alexa ecosystem mainly consists of two parts, customer home environment and Alexa cloud backend \cite{amazon2020aws}.
The home environment is the home setting of customers who have adopted
Alexa as their smart home extension, and typically contains Alexa smartphone app, Amazon Echo device, and other Internet-of-Things (IoT) devices.
Alexa cloud backend is comprised of Alexa voice service and supporting services from AWS (Amazon Web Services) cloud, e.g., Lambda serverless computing service, DynamdoDB database service, Amazon S3 storage service.
An overview of the Alexa ecosystem is presented in Fig.~\ref{fig:alexa-ecosystem} and the details of each component are provided below.

\begin{enumerate}
  \item Users: Alexa users can interact with Echo device via voice and other IoT devices.
  \item Echo: Echo device listens for the wake word and, once activated, it will record user voice and send recordings to Alexa voice service. When response (in text) is received, Echo device will play it in a lifelike voice.
  \item 3rd Party Alexa Devices: Devices made by third-parties (e.g., a raspberry pi running Alexa client). \item Alexa Voice Service: Alexa voice service performs smart SLU functionalities such as ASR and NLU, and TTS conversion in order to understand user request, which is used to decide which Alexa skill to invoke.
  \item Alexa Skill: Alexa skills are voice applications that respond user voice request. Alexa has built-in skills, e.g., providing weather forecasts, querying Wikipedia. New skills can be built to extend Alexa functionalities.
  \item Custom Skill API: A custom skill can be built with an AWS Lambda (a serverless computing service without provisioning servers) function that defines customized interactions with user request.
  \item DynamoDB with Auto Scaling: DynamoDB provides a NoSQL database that supports key-value and document based structure. With its high performance and scalability, DynamoDB can handle different usage scenarios. \item Alexa Smart Home Skill: Smart home skills allow users to control devices such as lights, thermostats, and other smart home devices via voice interaction.
  \item Smart Home Skill API: This API provides an interface for developers to describe the smart home devices and handle different user requests such as device discovery, status query and device control.
  \item AWS IoT: AWS IoT provides cloud services for connecting, monitoring and managing IoT devices in the cloud as well as services for analyzing device data.
  \item Amazon S3: Amazon S3 is an object storage service that can store static assets such as images, and media files corresponding to an Alexa skill.
  \item CloudFront: CloudFront provides a content delivery network service to serve content faster to Alexa users.
  \item IoT Devices: IoT devices are smart home devices that can be controlled with Alexa voice service.
  \item Audio Devices: Audio devices are the speakers that can play audio files such as music or recorded user voices.
\end{enumerate} \section{Attack Surfaces}
\label{sec:attack-surfaces}

\begin{figure}[htbp]
  \centerline{\includegraphics[width=0.5\textwidth]{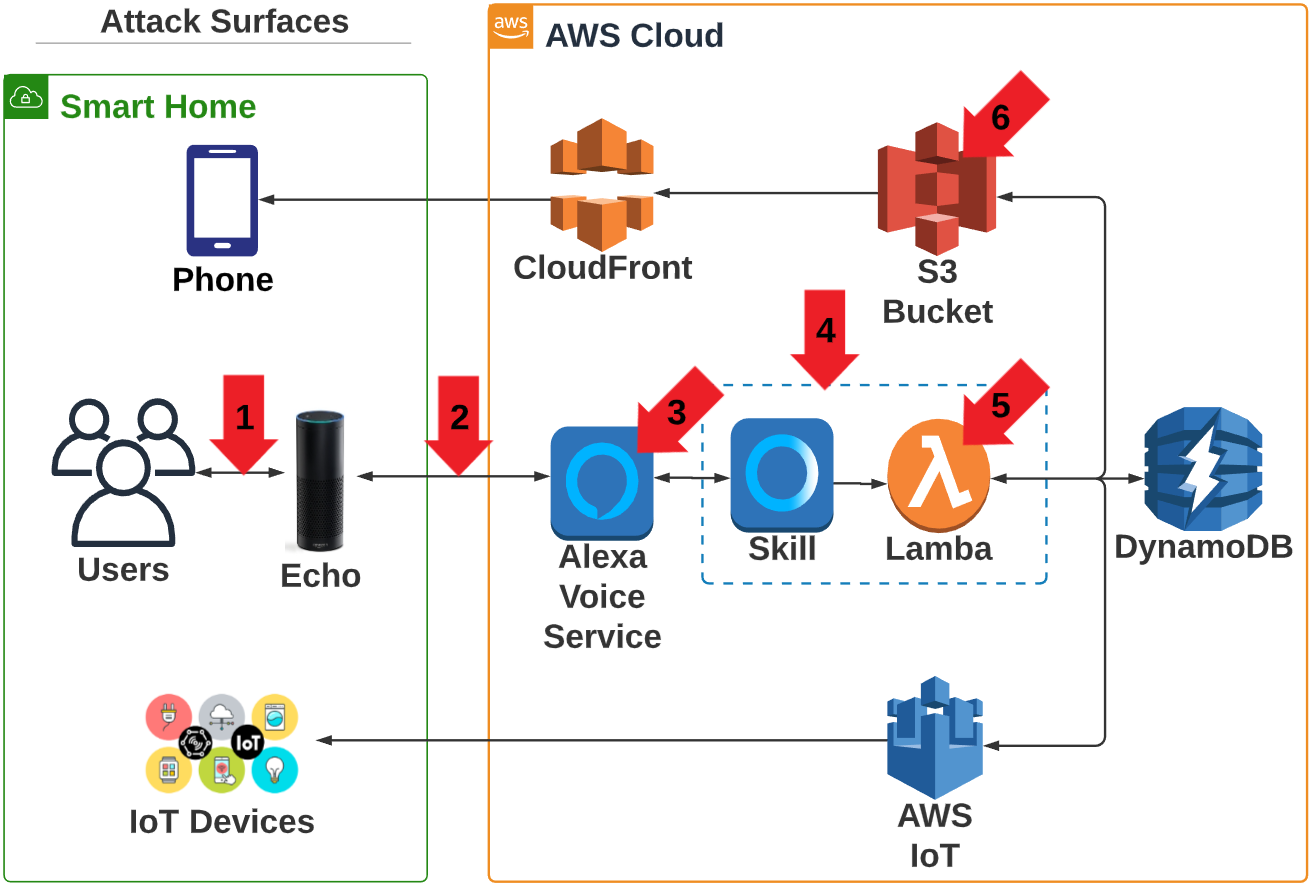}}
  \caption{Attacks surfaces of Alexa ecosystem}
  \label{fig:attack-surfaces}
\end{figure}

Amazon Alexa attack surfaces are identified by analyzing the existing attacks against Alexa ecosystem and categorizing which part of the Alexa ecosystem those attacks target. In total, we have identified six attack surfaces, depicted in Fig.~\ref{fig:attack-surfaces}.

\subsection{Voice Capturing (Attack Surface 1)}

The home environment is the most common setting for the Alexa voice assistant. With modern security measures, Alexa is able to prevent most noise pollution within this home environment from interrupting user conversations with the voice assistant. However, attacks that are able to permeate this noise pollution cancellation have been proven possible. These lingering security deficiencies make the voice capturing aspect of the Alexa a commonly targeted attack surface.

A very important aspect of the Amazon Alexa is its convenience. Once purchased, Amazon Echo device requires little set-up and can immediately be interacted with commands very similar to natural speech. Unfortunately, the Alexa lacks any form of voice-based authentication, allowing any voice within a home environment to interact and command Alexa. Therefore, any voice containing wake words can trigger Alexa. This has lead to the introduction of remote attacks, or attacks that take advantage of this lack of authentication by broadcasting commands over devices, such as television, radio, or speakers.

\textbf{Remote Voice Attack}:
Due to the lack of proper user voice authentication, voice commands played by any speakers can falsely trigger Echo device to respond. In \cite{yuan2018all}, researchers performed the remote voice attack in three different forms, injecting fake radio signals, replacing one TV channel with the provided video stream, and tricking a wireless speaker to play valid Echo commands.

\textbf{Dolphin Attack}:
Dolphin attack is another form of remote attack that instead uses inaudible commands to trigger Alexa. This inaudibility is possible by modulating voice commands on ultrasonic carriers. Although dolphin attacks require ultrasound transducers to be within 2 meters of the Echo device, making it a less common threat than the remote device attacks, there is still a concern of whether these sorts of attacks will be able to lengthen its attack distance in the future \cite{zhang2017dolphinattack}.

\textbf{Man-in-the-Middle}:
One research team was able to hack into an IoT device to attack the Alexa voice assistant. They used this along with several of the later mentioned techniques to implement a more sophisticated attack. Their “Man in the Middle” attack hijacks the conversation a user is having with their voice assistant without the user knowing. The first component of this attack is command jamming. They use the IoT device in order to inaudibly jam the commands the user is giving the voice assistant and simultaneously records them. When the user speaks the wake word, both the malicious and the voice assistant are activated. The malicious device uses an ultrasound modulated noise to prevent the voice assistant from understanding the user. The next component is data retrieval. Since the malicious device knows what skill the user was trying to use, it can send the same requests to the echo and find out the kind of data the user was looking for. The malicious device can now modify the data and complete the hijacking by echoing the information back to the user \cite{mitev2019alexa}.

\subsection{Voice Transmission (Attack Surface 2) }
In cybersecurity, fingerprinting refers to a set of information that can be used to identify network protocols, operating systems, hardware devices, software among other things. Hackers use fingerprinting as the first step of their attack to gather maximum information about targets.
The technique of fingerprinting can be used on the voice traffic between smart speakers and their cloud servers. In this attack, an adversary can eavesdrop both out-going and incoming encrypted voice traffic of a smart speaker, and infer which voice command a user says over encrypted traffic \cite{wang2020fingerprinting,ford2019alexa}.

\subsection{Alexa Voice Service (Attack Surface 3)}
These are attacks on the common SLU functionalities, ASR, NLU and TTS that Alexa voice service provides \cite{na2019alexa}.
At this day in age, it is nearly impossible for software to perfectly understand everything a person is saying and correctly interpret intent. Homonyms and homophones are commonly misunderstood even by humans, and has currently been found to be near impossible for computers to accurately identify in human language \cite{bispham2019nonsense}. Usually the language processing models are not accurate enough nor trained with other languages or accents. These deficiencies in Alexa's SLU are the targets for this particular attack surface.

\textbf{Skill Squatting}:
Skill Squatting is the creation of malicious skills that carry invocation and intent names that sound similar to the invocation and intent names of legitimate skills. Taking advantage of the easy misinterpretation of spoken word, skill squatting relies on the systematic errors produced from word-to-word, such as pauses or mispronunciations. The intent of this attack is to confuse the Alexa and empower the malicious skill to be used instead of the legitimate one, hijacking the legitimate skill. This technique can be focused on specific types of people by leveraging words that are only squattable in targeted users’ demographic \cite{kumar2018skill,zhang2019dangerous}.

\subsection{Alexa Voice Skill (Attack Surface 4)}
\textbf{Malicious skills}:
Malicious skills can be interpreted as any skill that is intentionally designed to act against the interests of the user. This could be a skill that mines user data or hijacks private information, or simply a skill that falsely claims to be able to perform an action it is unable to do. Currently, there are several studies that show how easily malicious skills are able to bypass Amazon's security and become available for public use on the official Amazon Alexa skill store \cite{leong2018analyzing,cheng2020dangerous,zhang2019dangerous}. In one study, researchers were able to publish hundreds of policy-violating skills onto the Alexa skill store for users to access. There proved to be many reasons why this attack surface in particular was so vulnerable \cite{cheng2020dangerous}.

In the current Amazon skill publishing system, once a user personally verifies that their skill follows Amazon policy, the skill is put up for review and screened by an Amazon official reviewer. If the official deems the skill as following customer policy, the skill becomes available for public consumption. In this way, while each skill is afforded the time and effort of an official review, the current Amazon skill publishing system is also very reliant on screening with little emphasis on objective automation. This subjective screening process leads to a dilemma of inconsistency. In other words, skill reviewers verify skills based on their own interpretation of the Amazon policy, meaning that the chance of a skill gets published is reliant not on the contents and intentions of the skill itself but of the skill reviewer assigned to its verification. This inconsistent verification process was demonstrated in a variety of different studies \cite{leong2018analyzing,cheng2020dangerous,zhang2019dangerous,barda2020checkpoint}.

In one specific example, researchers were also able to publish skills with malicious responses by delaying the response just enough to clear the process. Furthermore, if developers indicated (through a "developer's form" filled with yes or no questions) that their skill does not collect user information and data, they were granted publication onto the Amazon skill store, even if their indication was false, as that claim was often not verified by skill reviewers during the official review. This means that as well as relying on individual skill reviewers, Amazon is also reliant on individual developer honesty \cite{cheng2020dangerous}.

The research team also believed that the review process is largely manual due to the inconsistency in finding issues within skills which could have been easily found by automated systems. Additionally, there was evidence of many reviews being performed by non-native English speakers who may be unfamiliar with US laws \cite{cheng2020dangerous}. The study found that the review process was not thorough enough  to detect the malicious skills that had obvious policy-breaking functionalities. Through their research it can be suggested that the abundance of human reliance and lack of automation in the skill reviewing process is a security vulnerability \cite{cheng2020dangerous}.

\textbf{Masquerading Attack}:
Alexa voice skill is also vulnerable to voice masquerading attack. This attack takes place when the adversary makes a malicious skill that mimics the behavior of a legitimate skill or even the VPA service. These skills are able to convince the user that they are using safe and secure functionalities when in reality, all of the information they have given their voice assistance has been compromised \cite{zhang2019dangerous}.

\subsection{Lambda Functions (Attack Surface 5)}
With the adoption of serverless architecture comes the risk of applications being developed by insecure code. The OWASP community reported that these applications are vulnerable to traditional application-level attacks, like Cross-Site Scripting (XSS), Command/SQL Injection, Denial of Service (DoS), broken authentication and authorization and many more \cite{owasp2018github}.

\textbf{Command/SQL Injection}:
With the coming of voice assistants comes the voice user interface (VUI) and the VUI introduces new ways of transferring information that aren't secure or well monitored. This interface introduces a vulnerability into the Alexa system in the form of SQL injections. Attackers are able to use the VUI to interfere with queries to that the skills make to their databases and grant themselves access to otherwise sensitive data. In Black Hat Europe 2019, a group of researchers demonstrated that voice commands could also cause SQL injection if the Lambda function, processing user voice request, doesn't have proper input validation \cite{bannister2019port}.

\subsection{Amazon S3 Bucket (Attack Surface 6)}
Amazon Simple Storage Service (Amazon S3) is a scalable, high performance cloud storage service. An Amazon S3 bucket will be automatically created for storing media files when developers create an Alexa skill.

\textbf{S3 Bucket Misconfiguration}:
The access control to an S3 bucket could be misconfigured, e.g., configuring a bucket as publicly accessible to give skills (voice apps) an easy access to all the hosted files. The problem is, anyone could find the bucket by its name and get access to all files hosted in that bucket.
If that bucket happens to store some sensitive data such as private key or credentials, then those sensitive data would be leaked as well \cite{continella2018there}. Moreover, since websites could load resources from publicly writable S3 buckets, and if those buckets got maliciously modified, then the messages returned by Alexa skills could be fake, similar to Burger King example.

\section{Mitigation}
\label{sec:mitigations}

With the discovery of Alexa's attack surfaces, measures that minimize or eliminate these attack surfaces must be taken to ensure the safety of customer data. Unfortunately, the act of introducing additional security measures into a commercial product is already a very complicated issue. New features that enhance Alexa security must not only be effective at ensuring security; it must also perform in a way that does not disturb the convenient customer experience that has made Alexa so popular. Therefore, a healthy balance between security and usability must be found in order to optimize both customer safety and customer satisfaction.

\subsection{Mitigation to Voice Capturing Attacks}
One solution for protecting against remote voice capturing attacks is teaching Alexa to differentiate between live and recorded voices.
Void, proposed as a lightweight voice liveness detection system, is such an example \cite{ahmed2020void}. This software works to detect voice hacking attacks by finding the differences in spectral power (analysis of cumulative power patterns in spectrograms) between live-human voices and voices replayed through speakers through multiple deep learning models. Spectral power refers to the distribution of power into frequency components. Most loudspeakers inherently add distortions to original sounds while replaying them, making the overall power distribution over the audible frequency range show some uniformity and linearity.

Speaker-Sonar on the other hand is a sonar-based liveness detection system for smart speakers \cite{lee2020using}. Sonar is a technique that uses sound propagation to detect objects. The key idea for this system was to ensure that the voice command is indeed coming from the user by tracking user movement through a constant stream of inaudible ultrasonic sound and comparing the direction of the received voice command to the user's direction. This method in particular provided a non-intrusive user experience. However, it proved to only be reliably effectively in open outdoor spaces, as the highly decorated interiors of customer home environments proved to lessen the accuracy of the Speaker-Sonar system.

Protecting against Dolphin Attacks on the other hand requires a different set of solutions. A research team at Zhejiang University, the same research team that invented the Dolphin attack, introduced solutions for protecting against inaudible attacks. Hardware solutions for the inaudible attacks target the base of the problem. The root cause of dolphin attacks and other inaudible voice commands is that unfortunately, most commercial microphones attached to smart devices such as phones or voice assistants are able to detect acoustic sounds with frequencies higher than 20 kHz. Therefore, adjustments to microphones that suppress any acoustic signals whose frequencies are in the ultrasound range would effectively prevent many forms of inaudible attacks \cite{zhang2017dolphinattack}. Additionally, inaudible voice commands are able to be canceled by adding a module to microphones that detects modulated voice commands within the ultrasound frequency range. This module would then work to demodulate the signals to obtain the baseband \cite{zhang2017dolphinattack}.

\subsection{Mitigation to Voice Trasnmission Attacks}
Under an encrypted traffic analysis attack, an attacker can effectively eavesdrop on encrypted user interactions with smart speakers. In order to protect against the privacy leakage of smart speakers through voice traffic fingerprinting, a solution called ``adaptive padding" can be employed \cite{wang2020fingerprinting}. Adaptive padding refers to the addition of dummy packets to voice traffic, which are inserted based on the distribution of interarrival time while real packets still being sent at the original timestamp. This hides traffic bursts and traffic gaps, making encrypted user interactions with smart devices harder to interpret. Dummy packets produced by adaptive padding has the additional ability of sending buffered data sooner, effectively minimizing latency \cite{wang2020fingerprinting}.

\subsection{Mitigation to Alexa Voice Service Attacks}
Alexa voice service misinterpretations is one the most exploited attack surfaces. As previously mentioned in Section III, some common attacks that target this attack surface are voice squatting attacks, voice masquerading attacks, and skill squatting attacks.

A possible countermeasure against voice squatting and voice masquerading attacks is a skill-name scanner \cite{zhang2019dangerous}. The scanner would convert the invocation name string of a skill into a ARPABET-specified phonetic expression. This phonetic expression allows the phonetic distance between different skill names to be measured, and the skill names that are detected by the scanner to have a subset relation (considerable similarity) are deemed possible voice squatting attacks \cite{zhang2019dangerous}.
Another possible solution
is to take the context info, e.g., user's utterance and skill's response, into consideration. An example of this is presented in \cite{zhang2019dangerous}, in which the authors built a context-sensitive detector that consists of two major components, a user intention classifier and a skill response checker, to detect if there is an impersonation and give users an alert if impersonation is detected. This would ensure that the response of the skill match the perceived user intentions.

All skills, as previously discussed in Section III, must go through a certification process before they can be published to the Alexa skill store for public consumption. Skill squatting attacks rely on attackers successfully registering malicious squatted skills. In other words, skill squatting attacks rely on the flaws of this certification process. Therefore, a possible prevention tactic against skill squatting attacks is improving the certification process by adding additional screens. For instance, a word-based and phoneme-based analysis of a new skill's invocation name as a screening measure, in order to determine whether it could be confused for other already registered skills, would be an effective measure against skill squatting attacks \cite{kumar2018skill}.

\subsection{Mitigation to Alexa Voice Skill Attacks}
Voice assistant providers, such as Amazon, have certification processes that insufficiently check the skills submitted to their stores. A study in \cite{cheng2020dangerous}, provided two recommendations to help the providers to enhance the trustworthiness of their system. Since developers have the power to change the functionality of skills after their certification, enforcing the skill behavior integrity throughout the skill life-cycle is necessary. A continuous certification/vetting process should be required whenever the developer wants to change either the front-end or the back-end. Although this may increase the latency of the certification process, it will improve its quality and increase the trustworthiness of the system.

Another observation of the certification process for skill is the room for human error in a human decision dependent vetting process. An obvious fix would be to utilize automated skill testing in order to improve the consistency of the verifications and help the testing to be more thorough. According to \cite{cheng2020dangerous}, they concluded that 234 skill submissions were done in a largely manual manner and had limited voice response testing. To further increase the strength of the certification process, voice assistant system providers need to have access to the skill's back-end code to perform code analyses.

\subsection{Mitigation to Lambda Functions Attacks}
SQL injections have been around for a while and only because they still work and are effective at retrieving sensitive data. They are also effective at attacking the SQL databases used by Alexa skills if the Lambda functions are not up to par. One solution for that is to use Lambda-Proxy, which is an utility that could perform automated SQL injection testing for AWS Lambda functions \cite{segal2018security,segal2018puresec}. To better protect against Lambda based attacks, LambdaGuard, an AWS Lambda auditing tool could be used \cite{skyscanner2019medium}. LambdaGuard is designed to provide visibility into Lambda functions, conduct configuration checks to identify potential vulnerabilities.

\subsection{Mitigation to Amazon S3 Bucket Attacks}
Due to human errors for configuring S3 bucket, it is necessary to have stricter default policies and tools to check for common configuration errors. In \cite{continella2018there}, the authors developed a tool for bucket owners to check the access policies of their S3 buckets and verify if readable buckets contain sensitive data such as privacy keys in .pem file.
The same team also developed a browser extension for helping users to check if the rendered webpage loads resources from a writable S3 bucket \cite{continella2018there}. If so, the extension will prevent loading those untrusted resources. The same idea could be adopted by Alexa system to verify the legitimacy of the source. If the queried website from a skill is publicly editable or loads resources from a writable S3 bucket, then those should be blocked from playing.

\section{Conclusion}
\label{sec:conclusion}

Alexa voice assistant is a very popular consumer product
that has completely changed the way how people interact with smart
technology.
However, Alexa’s novel functionalities, such as its ability to
understand normal human conversation, adapt to the desires of the
customer, and control smart devices in the home environment,
require an unprecedented amount of user data, and unfortunately,
security and privacy has not been able to keep up to properly
protect that sensitive information. As a result, Alexa has been
the target of many attacks.

This paper surveyed and analyzed various attacks against Amazon
Alexa ecosystem, giving insights on where Alexa security risks
are located in its system. Furthermore, six attack surfaces were
identified by examining the lifecycle of Alexa voice interaction
that spans several stages including voice data collection,
transmission, processing and storage. In addition, mitigation
solutions to those attacks were also investigated and discussed
to provide directions for better improving Alexa or other voice
assistants in terms of security and privacy.

For future work, we plan to further evaluate the complexity of
each attack in the amount of efforts required from attacker's
perspective, as well as the amount of loss and the level of harm
it can bring to Alexa users from user perspective.

\bibliographystyle{IEEEtran} \balance

\end{document}